# AIDetection: A Generative AI Detection Tool for Educators Using Syntactic Matching of Common ASCII Characters As Potential 'AI Traces' Within Users' Internet Browser

v.1 (3/12/25)

Andy Buschmann[1]

**Abstract:** This paper introduces a simple JavaScript-based web application designed to assist educators in detecting AI-generated content in student essays and written assignments. Unlike existing AI detection tools that rely on obfuscated machine learning models, AIDetection.info employs a heuristic-based approach to identify common syntactic traces left by generative AI models, such as ChatGPT, Claude, Grok, DeepSeek, Gemini, Llama/Meta, Microsoft Copilot, Grammarly AI, and other text-generating models and wrapper applications. The tool scans documents in bulk for potential AI artifacts, as well as AI citations and acknowledgments, and provides a visual summary with downloadable Excel and CSV reports. This article details its methodology, functionalities, limitations, and applications within educational settings.

## 1. Motivation and Significance

The proliferation of generative AI tools has transformed the landscape of teaching and academic writing (Koivisto & Grassini, 2023; Lo, 2023; Rudolph, Tan, & Tan, 2023). While AI-assisted writing can enhance productivity, for instance, when used to proofread essays, it can also undermine critical and creative thinking (Bechky & Davis, 2025; Lindebaum & Fleming, 2024; Messeri & Crockett, 2024). Moreover, its use raises concerns regarding academic integrity and authorship (Rudolph et al., 2023; Thorp, 2023), including in educational settings where grades are often largely determined by writing assignments.

Despite the many unsettled questions and divergent opinions on how best to deal with ChatGPT and similar tools (Elbanna & Armstrong, 2024; Lau & Guo, 2023; Lim, Gunasekara, Pallant, Pallant, & Pechenkina, 2023), outright bans have little promise of success (Huang, 2023). In response, many instructors have resorted to stipulating policies around permitted use of AI, usually requiring students to provide appropriate citations and acknowledgments. To enforce such policies, however, educators need to be able to detect AI-generated content in student submissions and determine whether AI use has been acknowledged.

[1] University of Michigan, Ann Arbor, andybu@umich.edu

Many commercial AI detectors now exist that look for repetitive and 'unoriginal' ideas, as well as words and sentence structures that are 'typical' for generative AI (Kobak, González-Márquez, Horvát, & Lause, 2024). Not only are many of these tools 'black boxes' themselves, but their performance and reliability have been called into question (Khalil & Er, 2023; Messeri & Crockett, 2024). Ibrahim et al. (2023), for instance, evaluated two market-leading detection tools and concluded that: "AI-text classifiers cannot reliably detect ChatGPT's use in school work, due to both their propensity to classify human-written answers as AI-generated, as well as the relative ease with which AI-generated text can be edited to evade detection." Since the publication of their study and the release of even more advanced (reasoning) models, the problem has likely only worsened. Because of this, some schools, such as Vanderbilt University (Coley, 2023) and the University of Michigan (University of Michigan, 2025), advise against using AI detection software to generate 'proof' of cheating, while others have placed an outright ban on detection tools to prevent false accusations and perhaps costly future lawsuits (PLEASE, 2024).

While detection tools that analyze semantics may become increasingly unreliable as AI progresses, syntactic indicators of AI use are likely to persist. Attentive readers of student essays who have witnessed the pre- and post-ChatGPT era may have noticed that the formatting of certain characters is inconsistent within essays. A common reason is that students copy and paste passages directly from AI user interfaces into their text processors (such as MS Word or Google Docs). Most text processors use more-extensive text encodings such as UTF-8 and UTF-16, meaning they can display a larger variety of characters. Large Language Models (LLMs), by contrast, are usually trained on ASCII-formatted text because of greater compatability and smaller size (8 bit). More-widespread character encodings such as UTF-8 and UTF-16 are downward compatible with ASCII, meaning that ASCII chatacters are not automatically replaced. Hence, unless manually replaced by the author, these innocuous ASCII characters remain as potential 'AI traces' in the document and can be spotted.

While educators may visually detect these differences, they can be easy to miss, and visual examination can be tedious when grading long or numerous assignments. To simplify the task and employ additional heuristics, which I describe below, I created a simple JavaScript (JS)-based tool that is accessible online: aidetection.info. AIDetection allows educators to efficiently scan multiple documents—e.g., those downloaded in bulk from a course platform such as Canvas or Moodle—and assess the presence of AI-generated text as well as citations and acknowledgments. Since JS is executed within the user's own browser and no data is uploaded or processed on a web server, AIDetection complies with GDPR, FRAPA, and other data privacy regulations.

Before turning to its architecture, functions and advantages, I would like to put a disclaimer first: This tool does not replace contextualizing and sense-making by educators! Certain conditions must be met for this tool to be useful: First, the text needs to be written in a language that can be diplayed in ASCII (English). Languages that use different alphabets such as Chinese cannot be assessed. Second, generative AI applications must be widespread and used by those whose texts are being evaluated. (Among American college students today, this condition is reasonably met.) Third, only essays that were written after the adoption of ChatGPT and other generative AI tools has become widespread, can be assessed. This is particularly important because ASCII characters are not unique to AI; they could also appear when copying text from certain websites, such as Wikipedia. Pre-ChatGPT essays for which AIDetection could find matches may have engaged in plagiarism of human rather than AI content. Third, there must be inconsistencies in characters within the same text (which AIDetection already considers, as discussed below). Students may use operating systems such as FreeBSD or text editors that can only process ASCII characters. Lastly, esssays need to contain specific characters, which differ between ASCII and other text encodings. Hence, very short text (e.g., single sentences) are unlikely to yield results even if written by AI. Therefore, AIDetection is best used as a resource for educators to help monitor policy compliance or get an initial indication of possible AI (mis-)use, rather than as 'evidence' of cheating.

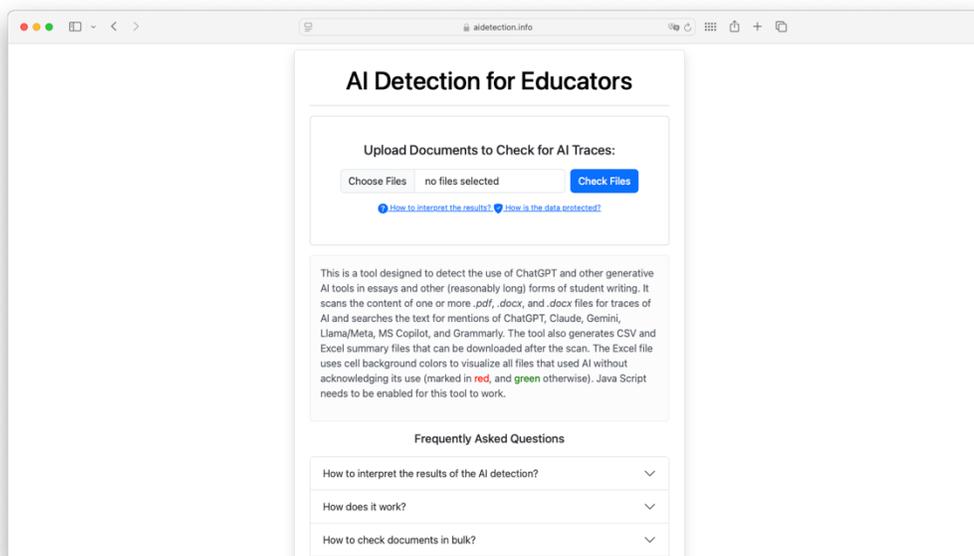

Figure 1: Screenshot of aidetection.info

**2. Software description**

**2.1. Software architecture and flow**

AIDetection.info relies on heuristic analysis rather than deep-learning classifiers. The tool detects AI-generated content by identifying inconsistent text encoding within text. Additionally, it scans for direct mentions of AI tools to determine whether students have acknowledged the use of AI. It consists of the following key components:

**File Processing Module:** The system begins by handling file uploads and determining whether a document should be analyzed based on its last modified date and file extension. Only PDF and Word files (.docx and .doc) are csupported. Files created before November 22, 2022, are ignored, as ChatGPT was not publicly available before this date. An error is returned for files that have an unsupported file extension as well as those files that were created prior to ChatGPT's launch.

**Text Extraction Module:** Extracting text from uploaded documents is essential for further analysis. The application utilizes Mozilla's PDF.js to parse text from PDF files and Mammoth.js to extract raw text from .doc and .docx files. These libraries are integrated via Cloudflare's content delivery network (cdnjs).

**Potential AI Trace Detection:** Once text is extracted, the system analyzes the content for syntactic artifacts and explicit mentions of AI tools such as ChatGPT, Claude, Grok, Gemini, Llama/Meta, Microsoft Copilot, and Grammarly AI. The detection process includes:
1. Identifying punctuation artifacts.
2. Detecting direct mentions of AI models and tools within the text.
3. Recognizing anomalous character encodings, i.e., a mix of ASCII and non-ASCII encodings for the same character.

**Results Processing and Visualization:** The detection results are displayed in an intuitive, color-coded summary:
- Green: No potential AI traces detected, or AI use was explicitly acknowledged.
- Red: Potential AI traces detected, but no explicit acknowledgment was found.

- Black: The document contains only ASCII-encoded text, and while potential AI traces are present, there is no regular human-formatted text available to make a conclusive determination.

**Export and Reporting Module:** Users can **download detection results** in multiple formats once all files have been processed. The available export options include:
- Excel (.xlsx) reports: Generated using the ExcelJS library (integrated via Cloudflare), allowing for sorting, filtering, and further analysis. Color coding is applied to indicate AI detection results.
- CSV (.csv) reports: Contain the same detection data but without cell-level color coding.

## 2.2. Potential AI Trace Detection

ASCII is the foundation for many modern encoding standards like UTF-8, which supports global languages while remaining downward-compatible with ASCII. AIDetection uses the latter feature as a heuristic to detect possible AI traces. To recognize anomalous character encodings, the application employs simple regular expression to the text and counts ASCII-encoded and non-ASCII single (apostrophe) and double quotation marks. (Regular expression is also used to detect the mentioning of common AI models.)

AI output for quotation marks and apostrophes in ASCII look like this: "straight double quotes" and 'straight single quotes/apostrophes,' while the output in Unicode (such as UTF-8) looks like this: "curly double quotes" and 'curly single quotes/apostrophes.' The differences are minimal and often hard to spot visually, especially in single-spaced documents, which is why using AIDetection can be helpful.

ASCII characters, particularly quotation marks, frequently appear in LLM outputs for multiple reasons. First, ASCII quotation marks are supported across all systems and do not rely on special character encoding, increasing compatibility. Second, AI models are usually trained on vast plain-text sources, which predominantly use ASCII-compatible characters. Relatedly, LLMs are also trained on published (and commented) source code. Programming languages are (usually) limited to ASCII characters. Third, AI models often optimize for 'tokenization' efficiency. ASCII quotation marks usually map to single tokens, whereas typographic quotation marks (e.g., curly quotes) require multi-byte encoding in formats like UTF-8. This increases token count and computational cost.

**2.3. Additional Website Stack**

I designed the aidection.info user interface using the Bootstrap framework, which is integrated using the jsDelivr content delivery network. In addition, I obfuscated the code on aidetection.info using JavaScript obfuscation (Kachalov, 2024) to prevent casual users, including students, to reverse engineer its matching method. The accompanying code consists of the un-obfuscated JS file as well as an unstylized HTML form.

**2.4. Software features**

AIDetection.info enables users to upload one or multiple .pdf, .docx, and .doc files at once. Key features include the potential AI trace detection (is based on counts of ASCII and non-ASCII quotation marks), the identification of AI acknowledgments through search of explicit references to ChatGPT, Claude, Gemini, Llama/Meta, Microsoft Copilot, and Grammarly (which can be easily expanded), the bulk processing of many files at once, and the visual representation.

Table 1: Columns in Downloadable Reports

| Column | Description |
| --- | --- |
| AI Traces | Total count of AI artifacts found in the document. |
| ChatGPT Mentioned | Whether any variant of ChatGPT was found in the document, either with or w/o whitespace, case-insensitive (e.g., "Chat gpt" or "CHATGPT"). If ChatGPT is cited or there is a statement of its use in the document, this will return Yes and No otherwise. |
| Grammarly Mentioned | Whether any variant of Grammarly was found in the document, case-insensitive. If mentioned, this will return Yes and No otherwise. |
| Claude Mentioned | Similar to the above. |
| Gemini Mentioned | Similar to the above. |
| Llama/Meta Mentioned | Whether any variant of Llama or Meta was found in the document, case-insensitive. If mentioned, this will return Yes and No otherwise. |
| Copilot Mentioned | Similar to the above. |

AIDetection generates downloadable Excel and CSV summaries with the columns in Table 1, using red highlights for unacknowledged AI use and green for acknowledged AI usage (for non-ASCII encoded text). See Figure 2 for an example. Lastly, the client-side execution ensures privacy by processing all files locally within the user's browser.

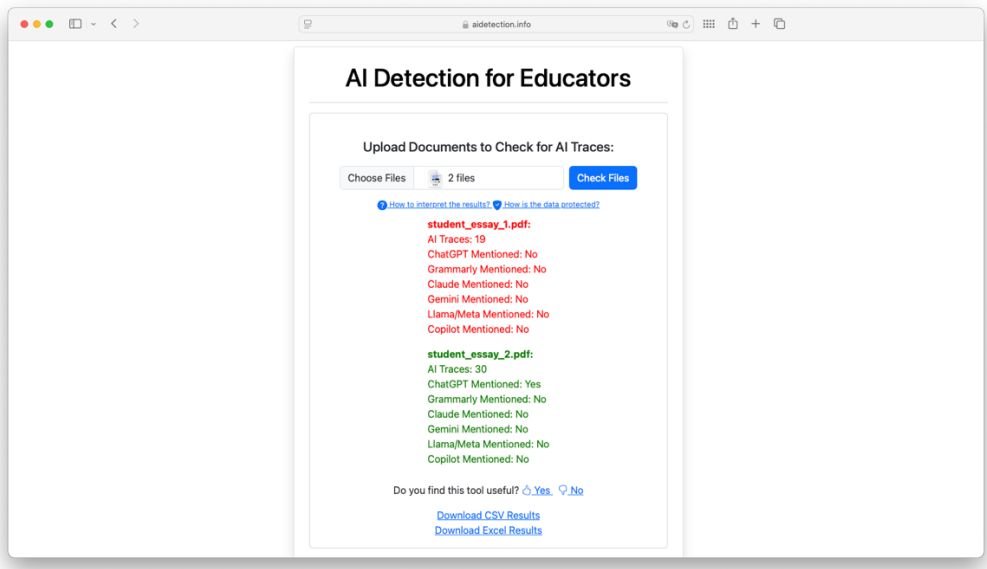

Figure 2: Screenshot of Results

The same color scheme depicted in Figure 2 applies to the Excel summary file (see Figure 3). In some instances, the results are ambigious. The reason is that the ASCII encodings are consistent throught the text. It may mean that the essay was fully written bei AI, but it may also be due to the text being written in an ASCII-based text editor. In those cases there is no color coding in the web UI. In the Excel summary file, such cases have a grey cell color.

Figure 3: Excel AI Detection Report

## 3. Illustrative example

I wrote this application while teaching an upper-level undergraduate writing class that required students to complete weekly writing tasks and four major essays. Despite having a clear AI policy that permitted AI use if properly acknowledged, students largely ignored this requirement. Stylistic inconsistencies and irregular character encodings were telltale signs of AI involvement.

I created AIDetection to monitor changes following a planned intervention. To implement this, I designed a survey presenting different AI usage scenarios. Students were required to indicate whether they believed acknowledgment was necessary in each case. The survey was mandatory and administered immediately after the submission of the second long-form essay.

Following the scenario-based questions, students who provided incorrect responses were shown the correct answers. Additionally, they were asked whether they had properly acknowledged AI use in their most recent essay. I emphasized that making corrections at this stage would not affect their grades if they were forthcoming. Students who had used AI without acknowledgment were then required to email me with details on what AI tools they had utilized.

As it turns out, all students flagged as having potentially used AI in their work and did not acknowledge it, had in fact used it. While a full analysis of the survey results is beyond the scope of this paper, it is noteworthy that most students struggled to apply the AI policy correctly, even in obvious cases where academic integrity should have been questioned. Additionally, tools such as Grammarly AI were frequently misidentified as simple text editors and thus not acknowledged. Similarly, Google's AI-generated summaries were often perceived as distinct from ChatGPT and not subject to the same disclosure requirements. For subsequent submissions, AI use was correctly acknowledged. (Since I did not seek IRB approval before the intervention, I am not reporting exact figures here.)

## 4. Impact

AIDetection.info provides educators with a simple-to-use and open-source tool to monitor AI policy compliance and measure the effectiveness of AI-related interventions. AIDetection supports checking individual assignments for AI-generated content as well as bulk-processing of submissions, which significantly reduces the workload for educators. By providing downloadable reports, the tool also provides a timestamped record when results are exported in Excel or CSV format. Lastly, because AIDetection.info does not exchange sensitive student data with external servers, ensuring compliance with FERPA and institutional data policies. It is free, quick, and easy to use, making it widely accessible to educators, and its transparent and non-black-box implementation provides full visibility into detection methods.

## 4.2. Scalability and Limitations

The client-side architecture makes the application efficient, but its performance dependents on the user's local computing resources. If a very large number of documents (>1000) were to be processed at once, internet browsers may stop responding or crash, depending on computing power. There are also more important limitations to consider: The tool does not provide definitive proof of AI-generated content; instead, it serves as a heuristic-based aid for educators. AI traces are 'potential AI traces' and as such, ASCII characters could originate from other copied sources like Wikipedia or uncommon document editors that do not support Unicode. Furthermore, it may yield false positives when analyzing essays discussing AI-related topics because names of common AI models are simply matched with in-text strings.

## 5. Conclusion

AIDetection.info provides a simple but useful solution for detecting potentially AI-generated content in student writing. By leveraging less-known heuristics, the application can be an effective and regulation-compliant alternative to no-detection software and sophisticated AI detectors. Its integration into AI interventions can help instructors convey the importance of citations and acknowledgments in academic writing, and support instructors in finding solutions to tackling issues associated with AI-assisted student writing.

## Bibliography

Bechky, B. A., & Davis, G. F. (2025). Resisting the Algorithmic Management of Science: Craft and Community After Generative AI. *Administrative science quarterly, 70*(1), 1-22. doi:10.1177/00018392241304403
Coley, M. (2023). Guidance on AI detection and why we're disabling Turnitin's AI detector. *Vanderbilt University*.
Elbanna, S., & Armstrong, L. (2024). Exploring the integration of ChatGPT in education: adapting for the future. *Management & Sustainability: An Arab Review, 3*(1), 16-29. doi:10.1108/MSAR-03-2023-0016

Huang, K. (2023). Alarmed by AI chatbots, universities start revamping how they teach. *The New York Times, 16*, e12255.

Ibrahim, H., Liu, F., Asim, R., Battu, B., Benabderrahmane, S., Alhafni, B., . . . Zaki, Y. (2023). Perception, performance, and detectability of conversational artificial intelligence across 32 university courses. *Scientific Reports, 13*(1), 12187. doi:10.1038/s41598-023-38964-3

Kachalov, T. (2024). javascript-obfuscator@4.1.1. Retrieved from https://github.com/javascript-obfuscator/javascript-obfuscator

Khalil, M., & Er, E. (2023). *Will chatgpt get you caught? rethinking of plagiarism detection.* Paper presented at the International conference on human-computer interaction.

Kobak, D., González-Márquez, R., Horvát, E.-Á., & Lause, J. (2024). Delving into ChatGPT usage in academic writing through excess vocabulary. *arXiv preprint arXiv:2406.07016*.

Koivisto, M., & Grassini, S. (2023). Best humans still outperform artificial intelligence in a creative divergent thinking task. *Scientific Reports, 13*(1), 13601.

Lau, S., & Guo, P. (2023). *From "Ban It Till We Understand It" to "Resistance is Futile": How University Programming Instructors Plan to Adapt as More Students Use AI Code Generation and Explanation Tools such as ChatGPT and GitHub Copilot.* Paper presented at the Proceedings of the 2023 ACM Conference on International Computing Education Research - Volume 1, Chicago, IL, USA. https://doi.org/10.1145/3568813.3600138

Lim, W. M., Gunasekara, A., Pallant, J. L., Pallant, J. I., & Pechenkina, E. (2023). Generative AI and the future of education: Ragnarök or reformation? A paradoxical perspective from management educators. *The International Journal of Management Education, 21*(2), 100790. doi:https://doi.org/10.1016/j.ijme.2023.100790

Lindebaum, D., & Fleming, P. (2024). ChatGPT Undermines Human Reflexivity, Scientific Responsibility and Responsible Management Research. *British Journal of Management, 35*(2), 566-575. doi:https://doi.org/10.1111/1467-8551.12781

Lo, C. K. (2023). What Is the Impact of ChatGPT on Education? A Rapid Review of the Literature. *Education Sciences, 13*(4), 410. Retrieved from https://www.mdpi.com/2227-7102/13/4/410

Messeri, L., & Crockett, M. (2024). Artificial intelligence and illusions of understanding in scientific research. *Nature, 627*(8002), 49-58.

PLEASE. (2024). Schools that Banned AI Detectors. Retrieved from https://www.pleasedu.org/resources/schools-that-banned-ai-detectors

Rudolph, J., Tan, S., & Tan, S. (2023). ChatGPT: Bullshit spewer or the end of traditional assessments in higher education? *Journal of applied learning and teaching, 6*(1), 342-363.

Thorp, H. H. (2023). ChatGPT is fun, but not an author. In (Vol. 379, pp. 313-313): American Association for the Advancement of Science.

University of Michigan. (2025). U-M Guidance for Faculty/Instructors. Retrieved from https://genai.umich.edu/resources/faculty